\documentclass[aps,multicol,psfig]{revtex4}
\newcommand{\beq}{\begin{equation}}
\newcommand{\eeq}{\end{equation}}
\newcommand{\bea}{\begin{eqnarray}}
\newcommand{\eea}{\end{eqnarray}}
\begin{document}

\title{
Interacting electrons in a one-dimensional
random array of scatterers \\
-- A Quantum Dynamics and Monte-Carlo study
}

\author{
V. Filinov$^{a,b}$, P. Thomas$^{b}$,
I. Varga$^{b,c}$, T. Meier$^{b}$,
 M. Bonitz$^{d}$, V. Fortov$^{a}$, and S.W. Koch$^{b}$,
}

\address{$^a$Institute for High Energy Density, Russian Academy of Sciences,\\
         Izhorskay 13/19, Moscow 127412, Russia\\
         $^b$Fachbereich Physik, Philipps-Universit\"at
         Marburg, D-35032 Marburg, Germany\\
         $^c$Elm\'eleti Fizika Tansz\'ek, Budapesti M\H uszaki \'es
         Gazdas\'agtudom\'anyi Egyetem, \\ H-1521 Budapest, Hungary\\
         $^d$Fachbereich Physik, Universit{\"a}t Rostock,
         Universit{\"a}tsplatz 3, D-18051 Rostock, Germany}
\date{\today}
\begin{abstract}
The quantum
dynamics of an ensemble of interacting electrons in an
array of random scatterers is treated using a
new numerical approach for the
calculation of average values of quantum operators and time correlation
functions in the Wigner representation.
The Fourier transform of the product of matrix elements of the dynamic
propagators obeys an integral Wigner-Liouville-type
equation. Initial conditions for this equation are given
by the Fourier transform of the Wiener path integral
representation of the matrix elements of the propagators at
the chosen initial times.
This approach combines both molecular
dynamics and Monte Carlo methods and computes numerical
traces and spectra of the relevant dynamical quantities such as
momentum-momentum correlation functions and spatial dispersions.
Considering as an application a system with fixed scatterers,
the
results clearly demonstrate that the many-particle interaction
between the electrons
leads to an enhancement of the conductivity and spatial dispersion
compared to the noninteracting case.
\end{abstract}
\maketitle

\section{Introduction}

Noninteracting electrons in an array of fixed
random scatterers are known to
experience Anderson localization at temperature $T=0$ in one-dimensional
systems. In two-dimensional systems single-parameter scaling theory
predicts that all states are localized as well \cite{gangoffour}. In
three-dimensional arrays of random scatterers localization appears at the
band extremities, while the center may consist of delocalized states,
indicating a disorder driven metal-insulator transition in three
dimensions only.

This picture has been challenged by recent experiments, which
suggest a metal-insulator transition also in disordered two-dimensional
electron systems \cite{sarachik}. A theoretical explanation \cite{dassarma}
indicated that the Coulomb interaction
between the electrons plays a central role in this effect.
Moreover, the experimental
study of persistent
currents in mesoscopic metal rings \cite{rings} yields currents that are
two orders of magnitude larger than predicted by theories
based on noninteracting electrons \cite{theory}. These findings
suggest that it is the many-particle interaction which leads to
delocalization tendencies. To the best of our knowledge
this has first been suggested by M. Pollak and coworkers
\cite{po1,po2}.

Evidence for this influence of the
Coulomb interaction has been obtained by examining the
problem of two interacting electrons in a one-dimensional
disordered band \cite{shepel}.
It is, however, not clear whether this schematic model
is able to yield answers for more realistic systems described by
an ensemble of many electrons. Therefore, attempts have been made to
calculate directly the conductivity of a disordered system of many
electrons. Numerical results for the ensemble averaged
logarithm of the conductance
have been obtained for
 small systems  by applying a
Hartree-Fock based diagonalization scheme for electrons without
\cite{epp1,epp2} and with spin \cite{epp3}.
An increase of the conductivity for certain regions in the
disorder-interaction parameter space has been demonstrated in these
publications.

The purpose of this paper is two-fold: i) We present our
approach, which does not rely on small system sizes.
This can be applied to a wide variety of different
physical systems, such as plasmas, liquids, and solids. ii)
As an illustration we study the problem described above.
In particular,
in this paper we investigate the influence
of the many-particle interaction on electronic transport.
We consider as a case study
a one-dimensional disordered array of scatterers interacting repulsively
with the electron system. Without electron-electron interaction
such a system shows Anderson localization. It is the purpose of this
application to study the effect of the long-range
electron-electron Coulomb interaction on the mobility of the electrons.

Anderson localization at temperature $T=0$ relies on quantum
coherence of electron trajectories and results from interference.
The key parameter in the physics of electron interference phenomena
is the dephasing time of electrons. At
finite temperatures the electron
coherence is destroyed
by both the
electron-electron and the phonon-electron scattering
\cite{ovtho,leeram,belitz,chaplik}.
To study the influence of these effects on  kinetic electron properties
in a random environment
we have simulated the quantum dynamics in a one-dimensional canonical
ensemble at finite
temperature for both interacting and noninteracting electrons
using a Quantum-Dynamics-Monte-Carlo scheme.
The main quantities calculated in this
paper are the temporal momentum-momentum
correlation functions,  their frequency-domain Fourier transforms and
the time dependence of the spatial dispersions.
We discovered that the results strongly depend  on the electron-electron
interaction, clearly
demonstrating the delocalizing influence of  the many-particle interaction
even at finite temperatures.

Our approach also treats the positions of the
scattering centers as dynamical variables.
We are therefore in the position to
generate various initial conditions, including the
scenario of ``electron bubbles'' as an interesting side-result.

We start in Section II
by presenting the Quantum-Dynamics-Monte-Carlo approach used in this work,
which is based of the Wigner representation of quantum statistical mechanics.
It treats both electrons and scatterers on an equal footing.
Details of our model, relevant correlation functions and the
system of reduced units
are introduced in Section III. Numerical
results for the momentum-momentum correlation function in the time
and frequency domains and for the spatial dispersion
are given in Section IV.
Finally, Section V contains a discussion of our results.

\section{Wigner representation of time correlation functions}

According to the Kubo formula
the conductivity is the Fourier transform of the current--current correlation
function.
Generally,
time correlation functions $C_{FA}(t)=\left\langle \hat{F}(0)\hat{A}%
(t)\right\rangle$ for a pair of dynamic variables $F(t)=\left\langle
\hat{F}(t)\right\rangle$ and
$A(t)=\left\langle
\hat{A}(t)\right\rangle$ are among the most
interesting quantities in the study
of the dynamics of electrons in disordered
systems of scatterers, such as
transport coefficients, chemical reaction rates,
equilibrium and transient spectroscopy, etc. Our
starting point is the general operator expression for the canonical
ensemble-averaged time correlation function \cite{zubar}:
\beq
C_{FA}(t) =Z^{-1}\mbox{Tr}\left\{\hat{F}\,e^{i\hat{H}t_c^{*}/\hbar}\,
\hat{A}\,e^{-i\hat{H}t_c/\hbar}\right\},
\label{cfa}
\eeq
where $\hat{H}$ is the Hamiltonian of the system expressed as a sum of the
kinetic energy operator, $\hat{K}$, and the potential energy operator,
$\hat{U}$. Time is taken to be a
complex quantity, $t_c=t-i\hbar \beta /2$, where
$\beta =1/k_BT$ is the inverse temperature with $k_B$ denoting the Boltzmann
constant.
The operators $\hat{F}$ and $\hat{A}$ are quantum operators of the dynamic
quantities under consideration and
$Z=\mbox{Tr}\left\{e^{-\beta\hat{H}}\right\}$ is the partition function.

The Wigner representation of the time correlation function in a
$\upsilon$--dimensional space can be written as:
\beq
C_{FA}(t)=(2\pi\hbar)^{-2\upsilon}
\int\int d\mu_1d\mu_2 \,F(\mu_1) \,A(\mu_2)\,W(\mu_1;\mu_2;t;i\hbar\beta),
\label{cfaw}
\eeq
where we introduce the short-hand notation for the phase space point,
$\mu_i=(p_i,q_i), (i=1,2) $, and $p$ and $q$ comprise the momenta and 
coordinates,
respectively,
of all particles in the system.
In the definition (\ref{cfaw}),
$W(\mu_1;\mu_2;t;i\hbar\beta)$
is the spectral density
expressed as
\bea
W(\mu_1;\mu_2;t;i\hbar\beta)&=&Z^{-1}\int\int d\xi_1d\xi_2 \,
e^{i\frac{p_1\xi_1}\hbar}e^{i\frac{p_2\xi _2}\hbar}\times \nonumber\\
&\times&\left\langle q_1+\frac{\xi_1}{2}\left|e^{i\hat{H}t_c^{*}/\hbar}\right|
                     q_2-\frac{\xi_2}{2}\right\rangle
        \left\langle q_2+\frac{\xi_2}{2}\left|e^{-i\hat{H}t_c/\hbar}\right|
                     q_1-\frac{\xi_1}{2}\right\rangle
,
\label{WW}
\eea
and $A(\mu)$ denotes Weyl's symbol \cite{tatr1} of the operator $\hat{A}$
\beq
A(\mu)=\int d\xi \, e^{-i\frac{p\xi}{\hbar}}
        \left\langle q-\frac{\xi}{2}\left |\hat{A}\right|
q+\frac{\xi}{2}\right\rangle,
\eeq
and similarly for the operator
$\hat{F}$.
Hence the problem of the numerical calculation of the canonically
averaged time correlation function is reduced to the computation of
the spectral density.

To obtain the integral equation for $W$ let us introduce the 
a pair of dynamic $p,q$-trajectories 
$\{\bar{q}_\tau (\tau ;p_1,q_1,t),\,\bar{p}_\tau (\tau ;p_1,q_1,t)\}$
and 
$\{\tilde{q}_\tau (\tau ;p_2,q_2,t),\,\tilde{p}_\tau (\tau ;p_2,q_2,t)\}$ 
starting at $\tau=t$ from the initial condition $\{q_1,p_1\}$ and $\{q_2,p_2\}$
propagating in `negative' and `positive' time direction, respectively: 
\bea
&&\frac{d\bar{p}}{d\tau}=\frac{1}{2}F\left[\bar{q}_\tau (\tau )\right]\,;
\qquad\qquad \frac{d\bar{q}}{d\tau}=\frac{\bar{p}_\tau (\tau )}{2m},
\nonumber \\
\mbox{with}\qquad && \bar{p}_t(\tau=t;p_1,q_1,t) = p_1\,;
\qquad \bar{q}_t(\tau=t;p_1,q_1,t)=q_1,
\label{a25}
\eea
and
\bea
&&\frac{d\tilde{p}}{d\tau}=-\frac{1}{2}F\left[\tilde{q}_\tau (\tau )\right]\,;
\qquad\qquad \frac{d\tilde{q}}{d\tau}=-\frac{\tilde{p}_\tau (\tau )}{2m},
\nonumber \\
\mbox{with}\qquad && \bar{p}_t(\tau=t;p_1,q_1,t) = p_1\,;
\qquad \tilde{q}_t(\tau=t;p_1,q_1,t)=q_1,
\label{aa25}
\eea

\beq
\mbox{while}\quad \varpi(s,q)=\frac 4{(2\pi\hbar)^\upsilon\hbar}
   \int dq^{\prime} \,
\tilde{U}(q-q^{\prime})\sin\left(\frac{2sq^{\prime}}{\hbar}\right)
         + F(q)\nabla\delta(s),
\eeq
$\delta(s)$ is the Dirac delta function and $F\left( q\right) \equiv - \nabla 
\tilde{U}$ with $\tilde {U}$ being the total potential.

Then as has been proved in \cite{filmd1,filmd2} the $W$ 
obeys the following integral equation:
\bea
W(\mu_1;\mu_2;t;i\hbar\beta)&=&
\bar{W}(\bar{p}_0,\bar{q}_0;\tilde{p}_0,\tilde{q}_0;i\hbar \beta )+
\nonumber\\
                            &+&\frac{1}{2}\int_0^td\tau\int ds \,
 W(\bar{p}_{\tau}-s,\bar{q}_{\tau};\tilde{p}_{\tau},\tilde{q}_{\tau};
   \tau ;i\hbar\beta) \, \varpi (s,\bar{q}_{\tau})-
\nonumber\\
                            &-&\frac{1}{2}\int_0^td\tau\int ds
\,
 W(\bar{p}_{\tau},\bar{q}_{\tau};\tilde{p}_{\tau}-s,\tilde{q}_{\tau};
   \tau ;i\hbar\beta) \,\varpi (s,\tilde{q}_{\tau}),
\label{a22}
\eea
where equation (\ref{a22}) has to be supplemented by an 
initial condition for the spectral density,
$\bar{W}(\mu_1;\mu_2;i\hbar\beta )\equiv W(\mu_1;\mu_2;t=0;i\hbar\beta)$,
which
can be expressed in the form of a finite difference approximation of the path
integral~\cite{filmd1,filmd2,feynm}:
\bea
\bar{W}\left( \mu_1;\mu_2;i\hbar \beta \right)
\approx
\int \int d\tilde{q}_{1} \dots d\tilde{q}_n
\int \int dq_{1}^{\prime} \dots dq_n^{\prime} \,
\Psi \left( \mu_1;\mu_2;\tilde{q}_1, \dots ,\tilde{q}
_n;
q_1^{\prime },\dots,q_n^{\prime};i\hbar \beta \right),
\eea

\bea
\mbox{with}
\quad \Psi \left( \mu_1;\mu_2;\tilde{q}_1,...,
\tilde{q}_n;q_1^{\prime
},...,q_n^{\prime };i\hbar \beta \right) \equiv
\nonumber\\
\frac{1}{Z}\left\langle q_1\left| e^{-\epsilon \hat{K}} \right|
\tilde{q}_1\right\rangle e^{-\epsilon U(\tilde{q}_1)}
\left\langle \tilde{q}_1\left| e^{-\epsilon \hat{K}}\right|
\tilde{q}_2\right\rangle
e^{-\epsilon U(\tilde{q}_2)} \:
\dots \:
e^{-\epsilon U( \tilde{q}_n)}\left\langle
\tilde{q}_n\left| e^{-\epsilon \hat{K}}\right|
q_2\right\rangle \varphi \left( p_2;\tilde{q}_n,q_1^{\prime }\right) \times
\nonumber\\
\left\langle q_2\left| e^{-\epsilon \hat{K}} \right|
q_1^{\prime }\right\rangle
 e^{-\epsilon U(q_1^{\prime})}
 \left\langle q_1^{\prime }\left|
e^{-\epsilon \hat{K}}\right| q_2^{\prime }\right\rangle
e^{-\epsilon U(q_2^{\prime})}
\dots e^{-\epsilon U(q_n^{\prime})} \left\langle
q_n^{\prime }\left| e^{-\epsilon \hat{K}} \right|
q_1\right\rangle \varphi \left( p_1;q_n^{\prime },\tilde{q}_1\right),
\label{a11a}
\\
\mbox{where}\quad
\varphi \left( p;q^{\prime },q^{\prime \prime }\right) \equiv
\left(2\lambda^2\right)^{\upsilon/2}
\exp \left[-\frac{1}{2\pi}\left\langle \frac{p\lambda}{\hbar}
+i\pi \frac{q^{\prime}-q^{\prime \prime}}{\lambda} \:\bigg |\:
\frac{p\lambda}{\hbar} +i\pi \frac{q^{\prime}-q^{\prime \prime}}{\lambda}
\right\rangle \right],
\label{a11}
\eea
where $\langle x |y\rangle$ denotes the scalar product of two vectors
$\vec{x}\cdot\vec{y}$.
In this expression the original (unknown) density matrix of the correlated
system $e^{-\beta(\hat{K}+\hat{U})}$
has been decomposed into $2n$ factors, each at a $2n$ times higher temperature,
with the inverse $\epsilon =\beta/2n$ and the corresponding
high temperature DeBroglie wave length squared
$\lambda^2\equiv 2\pi \hbar^2\epsilon/m$. This leads to a product of known
high-temperature (weakly correlated) density matrices, however, at the price of
$2n$
additional integrations over the intermediate coordinate vectors (over the
``path''). 
This representation is exact in the limit $n\rightarrow \infty$, and for
finite
$n$ an error of order $1/n$ occurs.

Expression (\ref{a11a}) has to be generalized to properly account for
spin effects.
This gives rise to an additional spin part of the density matrix,
whereas
 exchange effects can be accounted for by the antisymmetrization of only
one off-diagonal matrix element in (\ref{a11a}). As a result, in (\ref{a11a})
one matrix element, for example
$\left\langle q_1^{\prime }\left| e^{-\epsilon \hat{K}}\right| q_2^{\prime
}\right\rangle $ has to be
replaced by the antisymmetrized expression
$\left\langle q_1^{\prime }\left| e^{-\epsilon \hat{K}}\right| q_2^{\prime
}\right\rangle_A $
which can be written as the following sum of determinants involving
the {\em single-particle} density matrix
\bea
\left\langle q_1^{\prime }\left| e^{-\epsilon \hat{K}}
\right| q_2^{\prime }\right\rangle_A\equiv \sum_{s=0}^{N} C_N^s
\det\bigg|\left\langle q_{1,a}^{\prime }\left| e^{-\epsilon \hat{k}}
\right| q_{2,b}^{\prime }\right\rangle \bigg|_s = \sum_{s=0}^{N} C_N^s
\det\bigg|e^{-\frac{\pi}{\lambda^2}
\left|q_{1,a}^{\prime }-q_{2,b}^{\prime}\right|^2}\bigg|_s.
\label{det}
\eea
Here $C_N^s\equiv N!/s!(N-s)!$, and $|q_{1,a}^{\prime }-q_{2,b}^{\prime}|$
denotes the distance between the vertices $q_1^{\prime}$ and $q_2^{\prime}$
of particles $a$ and $b$, respectively.
As a result of the summation over spin variables and all possible exchange 
permutations, the determinant carries a subscript $s$ denoting the number of 
electrons having the same spin projection. To improve the accuracy of the 
obtained expression, in the total potential $U$ being the sum of all pair 
interactions $U_{ab}$, we will replace 
$U_{ab}\rightarrow U_{ab}^{\rm eff}$ where $U_{ab}^{\rm eff}$ is the
proper effective quantum pair potential, see below.
For more details on the path integral concept, we refer to Refs.
\cite{zamalin,prtthr,filinov76,znf76,filinov-etal.01ppcf},
for recent applications of this approach to correlated Coulomb systems, cf.
\cite{filinov-etal.00jetpl,filinov-etal.00jp,filinov-etal.01jetpl,afilinov-etal.01prl}.

Let us now come back to the integral equation (\ref{a22}). One readily checks
that its solution can be represented symbolically by an iteration series
of the form
\bea
W^t&=&\bar{W}^t+K_\tau ^tW^\tau \nonumber \\
   &=&\bar{W}^t+K_{\tau _1}^t\bar{W}^{\tau_1}
+K_{\tau_2}^tK_{\tau _1}^{\tau_2}\bar{W}^{\tau _1}
+K_{\tau_3}^tK_{\tau_2}^{\tau_3}K_{\tau_1}^{\tau_2}\bar{W}^{\tau_1}
+\dots,
\label{r3}
\eea
where $\bar{W}^t$ and $\bar{W}^{\tau _1}$ are the initial quantum spectral
densities
evolving classically during time intervals $[0,t]$ and $[0,\tau_1]$,
respectively,
whereas
$K_{\tau_i}^{\tau_{i+1}}$ are operators that govern the propagation
from time $\tau_i$ to $\tau_{i+1}$.

Since the time correlation functions (\ref{cfa}) are linear functionals of the
spectral density, for them the same series representation holds,
\bea
C_{FA}(t)=(2\pi\hbar)^{-2\upsilon}
\int\int d\mu_1d\mu_2 \,\phi(\mu_1;\mu_2)\,W(\mu_1;\mu_2;t;i\hbar\beta)
\equiv \left(\phi|W^t\right)
=
\nonumber\\
 \left(\phi|\bar{W}^t\right)
+\left(\phi|K_{\tau_1}^t\bar{W}^{\tau_1}\right)
+\left(\phi|K_{\tau _2}^tK_{\tau _1}^{\tau _2}\bar{W}^{\tau_1}\right)
+\left(\phi|K_{\tau _3}^tK_{\tau _2}^{\tau _3}K_{\tau _1}^{\tau_2}
       \bar{W}^{\tau_1}\right)
+\dots
\label{a31}
\eea
where $\phi(\mu_1;\mu_2)\equiv F(\mu_1)A(\mu_2)$ and the parentheses
$\left(\dots|\dots\right)$ denote integration over the phase spaces
$\{\mu_1;\mu_2\}$ as indicated in the first line of the equation.

Note that the mean value $\bar{F}(t)$ of a quantum operator $\hat{F}$
 can be represented in a form analogous to (\ref{a31}):
\beq
\bar{F}(t)=
(2\pi\hbar)^{-2\upsilon}
 \int\int d\mu_1d\mu_2\, \frac{F(\mu_1)+F(\mu_2)}{2}
\,
          W(\mu_1;\mu_2;t;i\hbar\beta),
\eeq
which allows us to increase the efficiency of the simulations. In the following
Section we apply this scheme to a system of interacting electrons and random
scatterers.

\section{Quantum dynamics}

As an application, in this work
we will consider a system composed of heavy
particles (called scatterers)
with mass $m_s$ and negatively charged electrons
with mass $m_e$.
To avoid bound state effects due to attraction we consider in this
case study only negatively charged scatterers, assuming a positve
backgroud for charge neutrality. The influence of
electron-scatterer attraction will be studied in a further
publication.

The possibility to convert a series like (\ref{a31}) into a form
convenient for probabilistic interpretation allows us to apply Monte Carlo 
methods to its evaluation. According to the general theory of the Monte Carlo 
methods for solving linear integral equations, e.g. 
\cite{sobol}, one can simultaneously calculate all terms of the iteration  
series (\ref{a31}). Using the basic ideas of \cite{sobol} we have 
developed a Monte Carlo scheme, which provides domain sampling of the 
terms giving the main contribution to the series (\ref{a31}), cf. 
\cite{filmd1,filmd2}. This sampling also reduces the numerical expenses for 
calculating the integrals in each term. For simplicity, in this work, we take 
into account only the first term of iteration series (\ref{a31}), which is 
related to the propagation of the initial quantum distribution (\ref{a11}) 
according to the Hamiltonian equation of motion. This term, however, does not 
describe pure classical dynamics but accounts for quantum effects 
\cite{filmd3} and, in fact, contains arbitrarily high powers of Planck's 
constant. The remaing terms of the iteration series describe momentum jumps 
\cite{filmd1,filmd2,filmd4} which account 
for higher-order corrections to the classical dynamics of the
quantum distribution (\ref{a11}) which are expected to be relevant in the limit
of high density. A detailed investigation of the conditions for which the
contribution of the next terms of the iteration series should be taken
into account is presented in \cite{filmd4,filmd5}.

The dynamical evolution of this system is studied along
the trajectory pair in phase space, Eqs.~(\ref{a25}) and (\ref{aa25}),
on time scales less than a maximum time $t^{\prime}$. 
This time is chosen small  enough such that
the system of scatterers is practically stationary within
the time interval $0 \le t < t^{\prime}$. The initial states of the system 
evolution are proper equilibrium states and thus relate
to physical times much larger than $t^{\prime}$. 
So the initial micro states (particle configurations in
phase space) can be randomly generated and an ensemble averaging should be performed 
with the canonical density operator of the electron-scatterer system, 
where the latter has been computed by the path integral Monte-Carlo method 
NEW \cite {zamalin} 
according to the probability distribution, which is proportional to 
$\left| \Psi  \right| $ (\ref{a11a}). 
So due to the time reversibility it is more convenient to start 
generation of the trajectories (Eqs.~(\ref{a25}) and (\ref{aa25}))  from time $\tau = 0$ starting 
from sampled by Monte Carlo method initial particle configurations (points $t_j$ on Fig.1).  
The phase cofactor of the complex-valued 
function $\Psi$ was taken into account by introducing the 
weight function \cite{sobol,Ceperley96,Berne98} of the initial configurations 
for the subsequent dynamic evolution. 
In principle, the method is also applicable to liquids or plasmas by
assigning smaller masses to the scatterers.

This approach allows us to generate, in a controlled way,
various kinds of initial conditions of the many-body system, in particular
(i) those which are
 characteristic of the fully interacting system [i.e.including
 scatterer-scatterer (s-s),
 electron-scatterer (e-s), and electron-electron (e-e)]
and (ii) those which result if
 some aspects of these interactions are ignored. In all cases,
the short-time dynamics can then be followed by including or excluding
the (e-e)-interaction.

For the numerical calculations we introduce dimensionless 
units, using the maximum value of the time characteristic for the short-time 
dynamics, $t^{\prime }$ as the unit of the dynamic time $t$ of the system. So 
the dimensionless time defined by $\theta =t/t^{\prime }$ 
will always vary between $0$ and  $1$. As a unit of length we take the 
reciprocal wavenumber $k^{-1}$, determined by the ratio 
$k^2=2m_e E_0/\hbar^2$. 
E.g., for one electron in an external potential field 
$\tilde{U}=V_0U(q)$ the operator exponent of the time evolution propagator 
$\exp(-i\hat{H}t/\hbar)$ can be rewritten in 
the form 
\beq
\frac{\hat{H}t}{\hbar}=
\left\{-\frac{\hbar ^2}{2m_e}\Delta +V_0U(q)\right\}\frac{t}{\hbar} =
\left\{-\frac{1}{2M}\Delta+\xi_0U(q)\right\} \theta
\eeq
where $\Delta$ is the Laplace operator, $V_0$ is characteristic strength of
the interaction potential, $M=\hbar/2E_0t^{\prime}$,
$\xi_0=V_0/2E_0M=V_0t^{\prime }/\hbar$.

The tensor of the electrical conductivity is given by:
\beq
\sigma_{\alpha\gamma}(\omega) = \int_0^\infty dt \,e^{i\omega t-\epsilon t}
  \int_0^\beta d\lambda
\,
  \left\langle\hat{J}_\gamma\hat{J}_\alpha(t+i\hbar\lambda)\right\rangle
\eeq
according to the quantum Kubo formula \cite{zubar},
where $\epsilon\rightarrow +0$. The current operator is
\beq
\hat{J}_\alpha=\sum_{i=1}^{N_e} e\dot{q}_i^{\alpha}(t)
              =\sum_{i=1}^{N_e} \frac{e}{m_e}p_i^{\alpha},
\eeq
with $\dot{q}^\alpha$ being the $\alpha-$component of the velocity operator of
the electron.
In the Wigner representation
this tensor reads
\beq
\sigma_{\alpha\gamma}(\omega)=\int_0^\infty dt \, e^{i\omega t-\epsilon t}
\int_0^\beta d\lambda \, \phi_{\alpha\gamma }(t,\lambda)\equiv
\frac{k^2}{e^2}\frac{\tilde{\sigma}_{\alpha\gamma}(\omega)}{2E_0t^{\prime}}
\eeq
where
\beq
\phi_{\alpha\gamma}(t,\lambda)=
\left\langle\hat{J}_\gamma\hat{J}_\alpha(t+i\hbar\lambda)\right\rangle=
\left (2\pi\hbar\right)^{-2\nu}\int d\mu_1d\mu_2 \,
J_\gamma(\mu_1)J_\alpha(\mu_2)W(\mu_1;\mu_2;t;i\hbar\beta;i\hbar\lambda),
\eeq
and the spectral density $W(\mu_1;\mu_2;t;i\hbar\beta;i\hbar\lambda)$
is defined as in Eq.~(\ref{WW}) by replacing $t_c^*$ by
$\tau_1=t_1+i\hbar\lambda$,
and $t_c$ by $\tau_2=t_2-i\hbar(\beta-\lambda)$.
The dimensionless conductivity tensor is denoted by
$\tilde{\sigma}_{\alpha\gamma }(\omega)$.

Our model of correlated electrons interacting with an array of random scatterers
is described by the following Hamiltonian,
\bea
\hat{H}&=&\hat{H}_{e}+\hat{H}_{es}+\hat{H}_{ss}
        =(\hat{K}_e+\tilde{U}_{ee})+\tilde{U}_{es}+(\hat{K}_s+\tilde{U}_{ss}),
\nonumber \\
\hat{H}_{e}&=&-\frac{1}{2M}\sum_{i=1}^{N_e}\Delta_i
              +\sum_{i\neq j}^{N_e}\xi_0^{ee}
                           u\left(\frac{|q_i-q_j|}{\lambda_{ee}}\right),
\nonumber \\
\hat{H}_{es}&=&\sum_{i=1}^{N_e}\sum_{j=1}^{N_s}\xi_0^{es}
                           u\left(\frac{|q_i-Q_j|}{\lambda_{es}}\right),
\nonumber \\
\hat{H}_{ss}&=&\frac{1}{2M}\frac{m_e}{m_s}\sum_{j=1}^{N_s}\Delta_j
              +\sum_{i\neq j}^{N_s}\xi_0^{ss}
                           u\left(\frac{|Q_i-Q_j|}{\lambda_{ss}}\right).
\eea
In the problem at hand we choose
$m_e/m_s \sim 1/2000\ll 1$, therefore, on the time scale
$t^{\prime}$ the scatterers are practically fixed. $\hat{H}_{ss}$
is the Hamiltonian of the scatterers, and
$\xi_0^{ee}=V_0^{ee}t^{\prime}/\hbar$,
$\xi_0^{es}=V_0^{es}t^{\prime}/\hbar$ and
$\xi _0^{ss}=V_0^{ss}t^{\prime}/\hbar$ are the constants of the binary
interaction
between electrons (e-e) $u\left(|q_i-q_j|/\lambda _{ee}\right)$,
electrons and scatterers (e-s) $u\left(|q_i-Q_j|/\lambda _{es}\right)$ and
between scatterers (s-s) $u\left(|Q_i-Q_j|/\lambda_{ss}\right)$,
respectively.
$q_i$ represent the positions of the electrons ($i=1,\dots,N_e$),
and $Q_j$ those of the scatterers ($j=1,\dots,N_s$).

The (e-e), (e-s) and (s-s) interactions are
taken in the form of the Kelbg potential~\cite{Ke63,kelbg}, which is the
exact quantum pair potential of an ensemble of weakly nonideal and weakly
degenerate
charged particles:
\beq
u(|x_{ab}|) =
\frac{1}{x_{ab}} \,\left[1-e^{-x_{ab}^2}+\sqrt\pi x_{ab}
\left(1-{\rm erf}(x_{ab})\right)
\right]
\label{kelbg-d}
\eeq
where $x_{ab}=|{\bf r}_{ab}|/\lambda_{ab}$,  with $\lambda$ being the
thermal wave length  given by
$\lambda^2_{ab}=\frac{\hbar^2\beta}{2\mu_{ab}}$, $\mu_{ab}$ the reduced mass
$\mu_{ab}^{-1}=m_a^{-1}+m_b^{-1}$,
$V_0^{ab}=\frac{e_a e_b}{\lambda_{ab}}$
and
 $e_a$, $e_b$ being the respective charges.
The ${\rm erf}(x)$ stands for the
error function. Note that the Kelbg potential is finite at zero distances and
coincides with the Coulomb potential at distances larger than $\lambda_{ab}$.
This potential has recently been successfully applied to the computation of
thermodynamic \cite{filinov-etal.00jetpl,filinov-etal.00jp,filinov-etal.01ppcf}
and
dynamic properties \cite{golub01pre} of dense quantum plasmas by means of path
integral
Monte Carlo and classical molecular dynamics methods, respectively. It is,
therefore,
expected to provide an accurate description of the full Coulomb interaction
between all particles within the quantum dynamics
approach of the present paper.

Due to the large difference in the masses  of 
electrons and heavy scatterers we can use 
two simplifications. The first one consists
in the antisymmetrization of the density
matrix only for electron spins and space coordinates.
The second one, as we mentioned above, is the use of the adiabatic
approximation for the dynamical evolution,
where the positions of the heavy scatterers in each initial
configuration  were fixed during the electron-dynamics time
given by the scale $t^{\prime}$.
The dynamic evolution has been realized according to
relations (\ref{a25}), (\ref{aa25}). Let us stress, that the Kelbg potential
appearing in the canonical density operator used for ensemble
averaging in (\ref{a11a})
has to be taken at the inverse temperature
$\epsilon \equiv \beta/2n$  while the simulation of
the dynamic evolution, according to Eqs.~(\ref{a25},\ref{aa25}) involves the
Kelbg
potential at the  temperature $1/\beta$.

To simplify the computations we included in (\ref{det}) only the dominant
contribution to
the sum over the total electron spin $s$ corresponding to $s=N/2$ electrons
having spin up and down, respectively. The contribution of the other terms
decreases
rapidly with particle number and vanishes in the thermodynamic limit. To further
speed up the
convergence of the
 numerical calculations we bounded the integration over the variables
$q_M,q_1^{\prime }, q_M^{\prime },\tilde{q}_1$
responsible for oscillations of the function $\varphi$ in (\ref{a11})
and checked the insensibility of the obtained results to these operations.

\section{Numerical results}

We now apply the numerical approach explained above to the problem of
an interacting
ensemble of electrons and disordered scatterers in one dimension.
In all calculations
and figures times, frequencies and distances are measured in atomic units.
The average distance
between electrons, $R_s=1/n_ea_0$, was  taken to be
$ 12.0; 2.6; 0.55$,
the densities of electrons and heavy scatterers are
equal.
The results obtained were practically insensitive to the variation of the
whole number of the particles in Monte Carlo cell from 30 up to 50 and
 also of the
number of high temperature density matrices
(determined by the number of factors $n$)  in (\ref{a11a}), ranging from
10 to 20. Estimates of the average statistical error gave the value of
the order  5-7 percent.
We studied two different temperatures:  $k_BT/|V_0^{es}|=0.45$ and  $ 0.28$,
corresponding to $\lambda_{ee}/a_0\sim 2.2$  and  $\lambda_{ee}/a_0\sim 3.5$,
respectively.
The strengths of the three interactions in the system are fixed
arbitrarily at the ratio
$V_0^{ee}:V_0^{es}:V_0^{ss} \sim 0.7:1:32 $.

According to the
Kubo formula \cite{zubar} our calculations include two different
stages: (i) the generation of the initial conditions
(configuration of scatterers and electrons) in the
canonical ensemble with probability proportional
to the quantum density matrix and (ii)
the generation of the dynamic trajectories on time scales
$t^{\prime}$ in phase space starting
from these initial
configurations.

The results presented below are related to three different situations.
The first two situations refer to the generation of initial configurations
used for ensemble averaging, where the (e-e) interaction is fully included
(called ``interacting ensemble''),
while the electron dynamics was simulated in
 case I. {\em with} (e-e) interaction (``interacting dynamics'') and, in case
II.,
{\em without} (e-e)  interactions (``noninteracting dynamics'').
 For reference a third situation, III.,
was studied where the (e-e) interaction is neglected completely, i.e. both, in
the initial ensembles (``noninteracting
ensemble'') and in the dynamical evolution, see
also Fig.~1 
for illustration. As we mentioned before to
avoid the influence of (e-s) bound state effects
we have considered only a system of
negatively charged heavy scatterers here.
The influence of (e-s) attraction on localization
will be studied in a further publication.

\subsection{Temporal quantum momentum-momentum correlation functions}

First we discuss the
influence of (e-e) interactions on the temporal quantum momentum-momentum
correlation function. 
Fig.~2 shows momentum-momentum correlation functions
corresponding to the situations I. and II. introduced above, i.e.
to interacting ensembles with interacting (curves 1)
and noninteracting (curves 2) dynamics,
respectively. 
Fig.~2 allows us to compare the
obtained functions for two temperatures and densities varying over one order
of magnitude.

The traces clearly show coherent oscillations, which are a
manifestation of localization tendencies even at finite
temperature. Note that at higher temperatures (right-hand panel)
these oscillations are
less pronounced compared to lower temperatures (left-hand panel).
The damping
times of these coherent oscillations are clearly different for the
noninteracting and interacting
dynamics. We observe that the electron-electron interaction
leads to a strong
reduction (curve 1) of this
damping time as compared to (e-e) noninteracting electrons (curve 2).
Furthermore,
for  the noninteracting dynamics
the appearance of deep aperiodic modulations  is seen.
The dynamics of (e-e) noninteracting electrons leads to at least one
large oscillation and several damped small ones, reflecting
a strong spatial confinement of the electron system in this case.
For interacting dynamics both, the first aperiodic modulation and
the subsequent damped coherent oscillations are less pronounced.

The damping time for coherent oscillations
increases with increasing density
and decreasing temperature. The physical reason of this phenomenon
is the tendency towards
ordering of the scatterers, which is obvious from the corresponding pair
distribution functions  ($g_{ss}$)
presented in Fig. 3, 
which develop periodic
modulations for higher densities.

As an interesting side-result 
Fig.~4 presents
the influence of the choice of initial conditions. In the left column we show
the momentum-momentum correlation functions
corresponding to situations II. and III., i.e. interacting (curves 1)
and noninteracting (curves 2) ensembles and, in both cases, dynamics without
electron-electron interaction.
The analysis of the equilibrium configurations and  pair distribution functions
for the noninteracting ensemble 
(right column of Fig.~4) shows that the main
contributions to the ensemble averaging  result  from
absolutely different particle configurations
if compared to the case of the interacting ensembles discussed
above. The
(e-e) noninteracting  electrons are gathered mainly in the extended deepest
potential well, while the heavy scatterers form a potential barrier
confining the electrons. The typical initial configuration for
equilibrium averaging, therefore, contains a so called `electron bubble'.
The (e-e) noninteracting electrons
oscillate inside this `bubble'
with frequencies defined by the eigenvalues of the
resulting potential energy profile
of this well. On the other hand, for the typical configurations of
interacting ensembles (cf. Fig. 3) 
the scatterers
and electrons are both distributed more or less uniformly in space and
the potential profile shows only spatial small-scale random inhomogeneities.

In the case of noninteracting ensembles we can see
high frequency coherent oscillations  with
frequencies defined by the  potential forming the
`bubble', while for interacting ensembles
with their spatial small-scale potential profile
only aperiodic oscillations (cf. Fig. 2, 
curves 2) are seen.

\subsection{Fourier transform of the momentum--momentum time correlation
functions}

Figures 
5 - 7  present the real and imaginary parts
of the diagonal elements of the electrical conductivity tensor versus frequency,
i.e. the real and imaginary parts of the
Fourier transform of the temporal momentum--momentum correlation functions.
The real part of the Fourier transform characterizes the Ohmic absorption of
electromagnetic energy and has the physical meaning of electron
conductivity, while the imaginary part presents $(\epsilon -1)\omega $,
where $\epsilon $ is the permittivity of the system.

Curves 1 and 2 of 
Fig.~5  show an opposite behavior for vanishing frequency.
Most remarkably, the low-frequency conductivity related
to the interacting dynamics (curve 1)
is positive, while that for the noninteracting
dynamics (curve 2) has a maximum at some finite frequency related
to the coherent oscillations in the time domain. For lower
frequency it changes sign. However, the
real part of the conductivity has to be nonzero. The negative
contributions are due to weakly
damped very slow oscillations with time scales exceeding the scale
$t^{\prime}$ considered for the calculation of the dynamics
for the noninteracting dynamics
(see the temporal momentum-momentum correlation functions, 
Fig.~2).
To overcome this deficiency of our
model one has to increase the time $t^{\prime}$
and/or to take into account the slow motion of the heavy
particles, which will destroy the coherent oscillation of the light
electrons trapped by the heavy particles and
thus suppress these negative values. In fact, we found that
calculations performed
for longer and longer times lead to decreasing negative
contributions for low frequencies.

Nevertheless, let us stress that the results obtained 
(Figs. 2 and 5) allow us to conclude that the behavior of the conductivity
in the vicinity of zero frequency resembles the characteristic
features of Anderson localization (i.e. a vanishing
zero-frequency conductivity and enhanced
oscillations related to the maximum in the real part
at finite frequency) for noninteracting dynamics
of electrons. The
(e-e) interaction, on the other hand, leads to a strong increase of the
conductivity at low frequencies and less pronounced oscillations.

Let us note that the high frequency tails of
the real and imaginary parts of the Fourier transforms presented in
Figs.~5 and 6
coincide with each other, as the main contribution is due to
the fast trajectories with high virtual energy.
This means, the (e-e) interaction does not influence the behavior of the high
frequency tails.
Furthermore, we have checked that the high-
frequency tails of the real and imaginary parts
of the Fourier transform can be
described by the Drude-asymptotics for free electrons. This reflects the fact
that also the (e-s) interaction (its coupling constant
being comparable to that of the
(e-e) interaction) does not effect the
behavior of the high-energy trajectories.

The Fourier transform of the correlation functions
for noninteracting dynamics and both interacting and noninteracting
ensembles
are presented in 
Fig.~7 (cf. also Fig.~4). The most
remarkable
feature of these figures is the splitting of the peak related
to the coherent electron oscillations for the noninteracting
ensembles. The positions of
the inserted triangles show the energy difference between
subsequent eigenenergy levels calculated for an averaged bubble.
The comparison of the peak and the triangle
positions confirms the fact that
\cite{zubar} the
Fourier transform of the momentum-momentum correlation function
contains a sum of products of delta-functions and momentum matrix
elements related to transitions between these energy levels.
The imaginary part of the Fourier transform presented in 
the right column of Fig.~7 reflects that these characteristic features are
related to the
resonance oscillations in the `bubble' mentioned above.

Figure 
5 shows that for noninteracting dynamics,
at low temperatures and moderate densities, e.g. $R_s=2.6$
(cf. Figs.~ b), a well-resolved
absorption peak appears which is related to the electron energy level
separation.
Note that this peak is present even in the case of interacting ensembles.
However, there, the height of this peak is considerably
smaller than in the case of noninteracting ensembles 
(Fig.~7).

\subsection{Position dispersion}

Finally, we discuss the position dispersion of electrons for
interacting and noninteracting ensembles and noninteracting dynamics, cf.
Fig.~8. Electron localization in the bubble results in a very slow
growth (curve 2) of the position dispersion in comparison with the case of
the small-scale potential profile (curve 1) of the initial configurations,
as seen in Fig.~8. 
The square root of the position dispersion at times 
of the order of unity
yields the typical size of the bubble.
However, for larger times, $t>80$, curve 2 shows a behavior, which is
typical for particle diffusion. This characteristic change in behavior is
due to the tail of the electron energy distribution function
and is related to the fact that we are dealing with non-selfaveraging
quantities.
For the high-frequency tails of the Fourier transforms the
virtual energy of the trajectories may be large. These exponentially
rare fast trajectories can give
an exponentially large contribution to the position dispersion, as the
difference in positions of fast trajectories may be exponentially large at
sufficiently large time. A similar problem connected with exponentially
large contributions of exponentially rare events arises in the
consideration of classical wave propagation in random media. There, it is well
known \cite{gred1} that
in one-dimensional systems the dispersion of wave intensities
is not self-averaging, as exponentially rare configurations of
scatterers can give rise to exponentially large contributions of intensity at
large distances from the wave source.

\section{Conclusion}
The Quantum-Dynamics-Monte-Carlo
approach applied to a one-dimensional system
of interacting electrons in an array of fixed random scatterers
at finite temperature gives strong evidence for an enhancement
of the quantum mobility of electrons due to their mutual long-range
many-particle
interaction
and thus substantiates previous expectations drawn from
schematic models.

As in our approach
the temperature is taken to be nonzero, Anderson localization
is not expected to show up in the results for
noninteracting electrons in a strict sense. On the other
hand, for the considered temperatures (comparable to or less than the coupling
constant of the electron-scatterer potential) localization tendencies are
clearly observable in this case. They
manifest themselves both,
in the low-frequency behavior of the momentum-momentum correlation function
related to the conductivity of the electron system, and in
coherent oscillations. For nonzero electron-electron
interaction these localization tendencies are relaxed and the frequency-
dependent
correlation function has a form not unlike a Drude behavior.
The high-frequency tails resemble a Drude-behavior of free
electrons in all cases.

For initial conditions describing a noninteracting ensemble,
the scattering centers form electron-bubbles which
accomodate the [(e-e) noninteracting] electrons in their eigenstates and lead
to well-defined coherent oscillations.
In contrast, for configurations
describing interacting ensembles, the scatterers
and electrons are both distributed more or less uniformly in space and
the potential profile shows only spatial small-scale random inhomogeneities.
Then the dynamics of interacting electrons
in this spatial small-scale random potential profile shows
only fast damped oscillations with much smaller damping time if
compared to noninteracting dynamics
and results in an
increased low-frequency conductivity.

The present work shows that it is possible to treat a system of many
quantum particles interacting with each other via the long-range Coulomb
potential in a numerically
rigorous scheme. Here we have considered the heavy particles
to be essentially immobile species. On the other hand, if their mass is being
reduced
they will take part in the dynamics, and we then have a system resembling
a plasma of, e.g., electrons and holes. Such calculations are
presently under way and will be published shortly.

\section{Acknowledgments}
The authors thank B.L. Altshuler for stimulating discussions and valuable
notes and the NIC J\"ulich for
computer time. V. Filinov acknowledges the hospitality
of the Graduate College
``Optoelectronics of Mesoscopic Semiconductors'' and the Department
of Physics of the Philipps-Universit\"at Marburg.  This work is
partly supported by the Max-Planck Research Prize of the 
Humboldt and Max-Planck Societies, by the Deutsche Forschungsgemeinschaft
(DFG) through the Quantenkoh\"arenz Schwerpunkt, and by the
Leibniz Prize. I.V. acknowledges financial support from the
Hungarian Research Fund (OTKA) under T029813, T032116 and T034832.

\newpage

\ \ \ \ \ \ \ \ \ \ \ \ \ \ \ \ \ \ \ \ \ \ \ \ \ \ \ \ \ \ \ \ \  FIGURES \\ \\ 

Fig.1 Illustration of the Quantum-Dynamics scheme for computation of the
correlation function $C_{FA}(\tau)$.
The quantum general dynamics scheme generates particle trajectories (long arrows) in
phase space $p-q$ starting
from an initial state at $t=0$. The two arrows indicate different ensembles used
for averaging over the initial states which do
(do not) include e-e interaction: interacting
ensemble, (noninteracting ensemble).
At randomly chosen times $t_1, \dots, t_i, t_{i+1},\dots,t_K$,
a pair of dynamic trajectories (short arrows)
is propagated in positive and negative
time direction, e.g. from $t_i$ to $t_i+\tau/2$ and  $t_i-\tau/2$, respectively.
$C_{FA}(\tau)$
is averaged over all $K$ trajectories. Bold (dashed) arrows indicate
(non-)interacting dynamics including (neglecting) e-e interaction.

Fig.2 Temporal momentum-momentum correlation for interacting ensembles for the case of
interacting (1) and noninteracting (2) dynamics. Temperatures are
$k_BT/|V_0^{es}|=0.28$
(left panel: a,b)
and  $k_BT/|V_0^{es}|=0.45$ (right panel: c,d).
 
Fig.3 e-e, e-s and s-s pair distribution functions for the interacting ensemble,
for the same parameters as in Fig. 2.

Fig.4 Left Column: Temporal momentum-momentum correlation functions for the case of 
noninteracting
dynamics and interacting (1) and noninteracting (2) ensembles. Temperature is
$k_BT/|V_0^{es}|=0.45$.
Right Column: Pair distribution functions for the noninteracting ensembles.  

Fig.5 Real part of the Fourier transform of the temporal
momentum-momentum correlation functions
of Fig. 2 

Fig.6 Imaginary part of the Fourier transform of the temporal
momentum-momentum correlation functions
of Fig. 2. 

Fig.7 Real part (left column) and imaginary part (right column) of the Fourier
transform of the temporal momentum-momentum
correlation functions of Fig. 4. 
Triangles indicate the transition energies between
the lowest energy eigenvalues for an anlytical model for the deepest
scatterer potential
which traps the electrons
(vertical coordinates of the triangles are arbitrary).

Fig.8 Electron position dispersion for noninteracting
dynamics and interacting (1) versus noninteracting  (2)
ensembles, cf. Fig. 4.

\newpage

\begin{figure}[f0]
\vspace{-2.cm}
\caption[]{\label{f0}
Illustration of the Quantum-Dynamics scheme for computation of the
correlation function $C_{FA}(\tau)$.
The quantum dynamics scheme generates particle trajectories (long arrows) in
phase space $p-q$ starting
from an initial state at $t=0$. The two arrows indicate different ensembles used
for
averaging over the initial states which do
(do not) include e-e interaction: interacting
ensemble, (noninteracting ensemble).
At randomly chosen times $t_1, \dots, t_i, t_{i+1},\dots,t_K$,
a pair of dynamic trajectories (short arrows)
is propagated in positive and negative
time direction, e.g. from $t_i$ to $t_i+\tau/2$ and  $t_i-\tau/2$, respectively.
$C_{FA}(\tau)$
is averaged over all $K$ trajectories. Bold (dashed) arrows indicate
(non-)interacting dynamics including (neglecting) e-e interaction.
}
\end{figure}
\begin{figure}[f1]
\vspace{-0.5cm}
\caption[]{\label{f1}
Temporal momentum-momentum correlation for interacting ensembles for the case of
interacting (1) and noninteracting (2) dynamics. Temperatures are
$k_BT/|V_0^{es}|=0.28$
(left panel: a,b)
and  $k_BT/|V_0^{es}|=0.45$ (right panel: c,d).
}
\end{figure}

\begin{figure}[f2]
\vspace{0.5cm}
\caption[]{\label{f2}
e-e, e-s and s-s pair distribution functions for the interacting ensemble,
for the same parameters as in Fig. 2.
}
\end{figure}

\begin{figure}[f3]
\vspace{-4.5cm}
\caption[]{\label{f3}
Left Column: Temporal momentum-momentum correlation functions for the case of 
noninteracting
dynamics and interacting (1) and noninteracting (2) ensembles. Temperature is
$k_BT/|V_0^{es}|=0.45$.
Right Column: Pair distribution functions for the noninteracting ensembles.  
}
\end{figure}


\begin{figure}[f5]
\vspace{-0.2cm}
\caption[]{\label{f5}
Real part of the Fourier transform of the temporal
momentum-momentum correlation functions
of Fig. 2 
}
\end{figure}

\begin{figure}[f6]
\vspace{-0.2cm}
\caption[]{\label{f6}
Imaginary part of the Fourier transform of the temporal
momentum-momentum correlation functions
of Fig. 2. 
}
\end{figure}

\begin{figure}[f7]
\vspace{-5cm}
\caption[]{\label{f7}
Real part (left column) and imaginary part (right column) of the Fourier
transform of the temporal momentum-momentum
correlation functions of Fig. 4. 
Triangles indicate the transition energies between
the lowest energy eigenvalues for an anlytical model for the deepest
scatterer potential
which traps the electrons
(vertical coordinates of the triangles are arbitrary).
}
\end{figure}


\begin{figure}[f9]
\vspace{-5.5cm}
\caption[]{\label{f9}
Electron position dispersion for noninteracting
dynamics and interacting (1) versus noninteracting  (2)
ensembles, cf. Fig. 4.}
\end{figure}

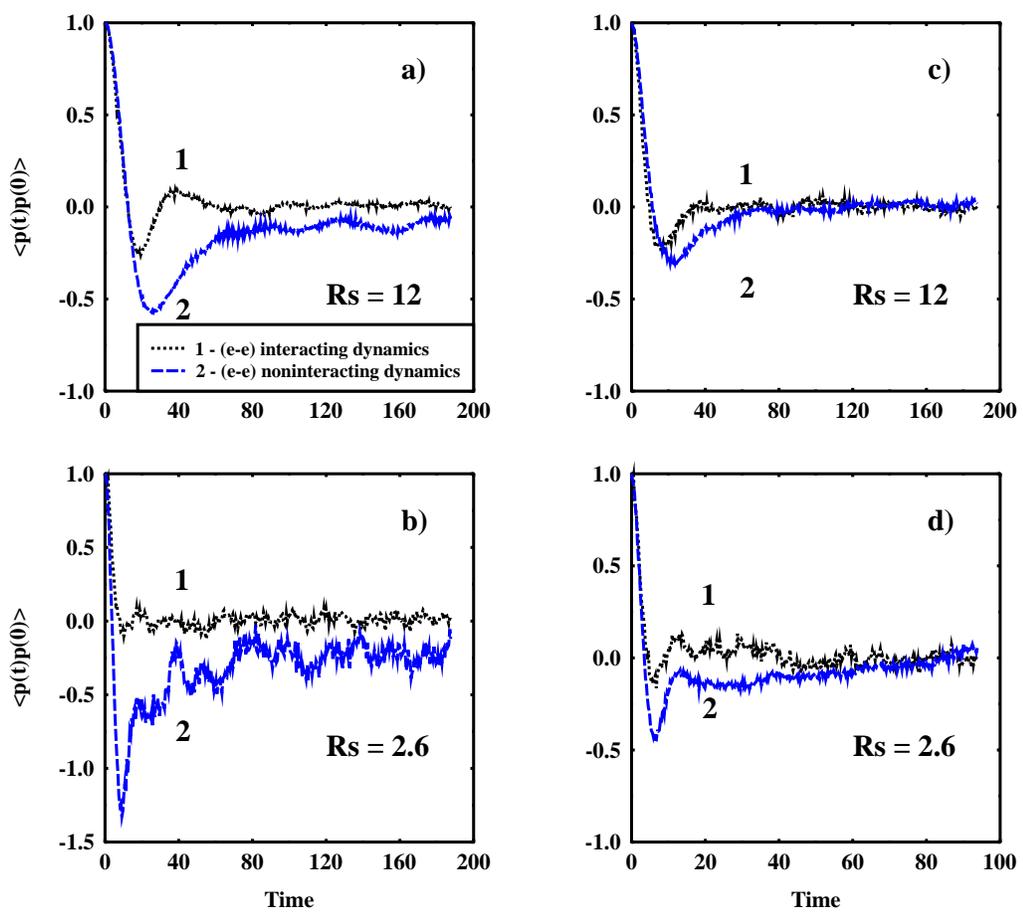
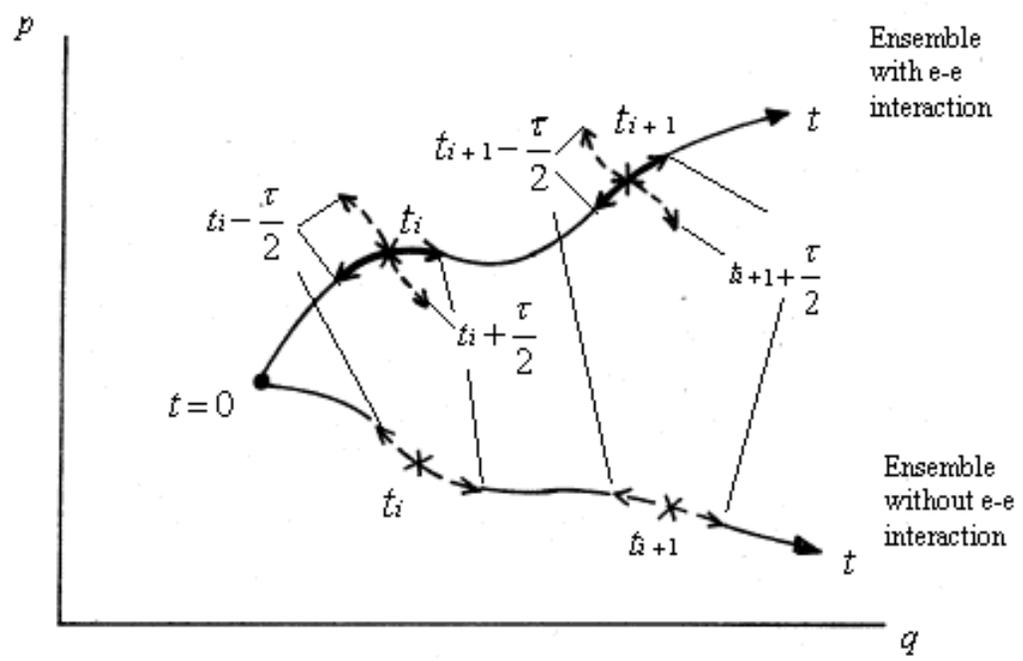
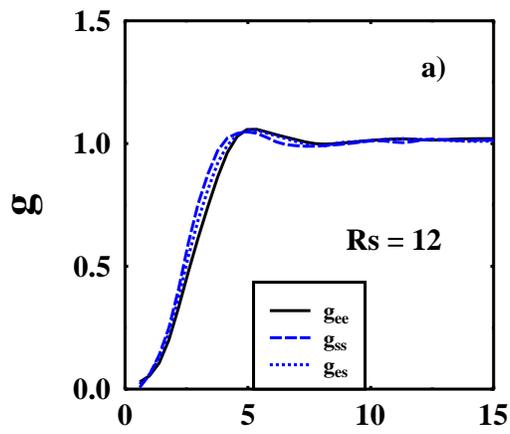
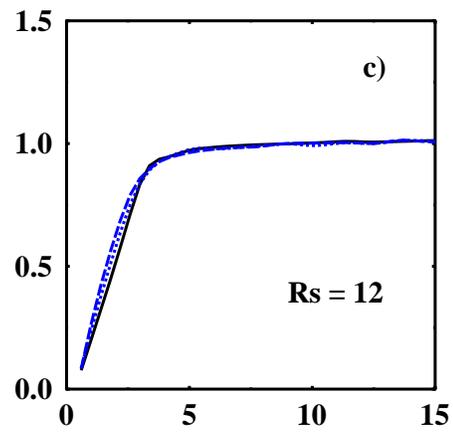
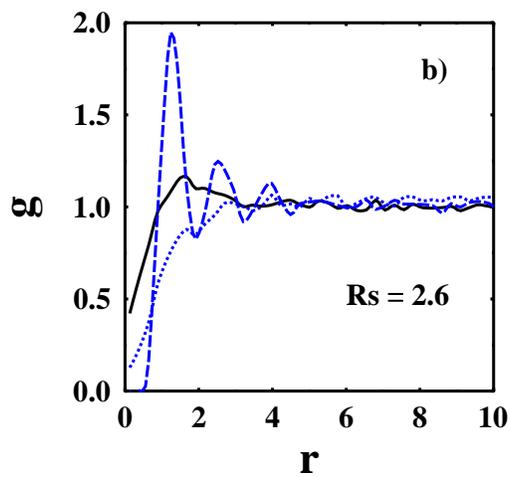
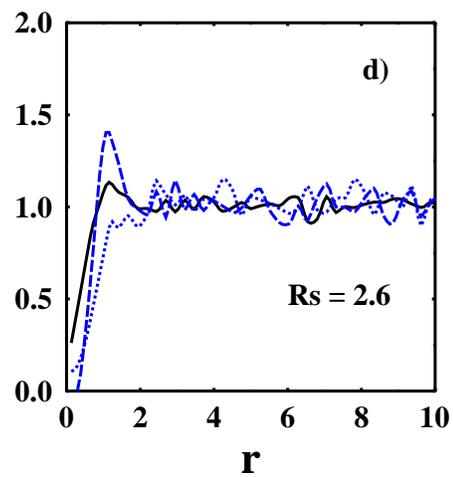
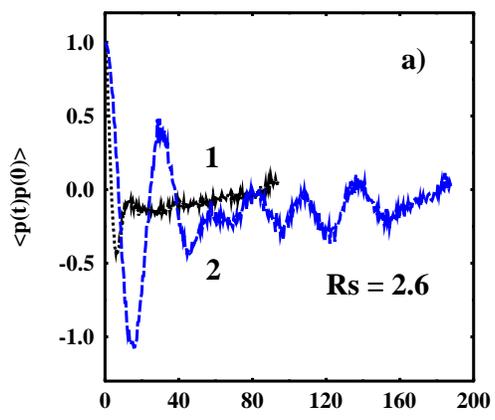
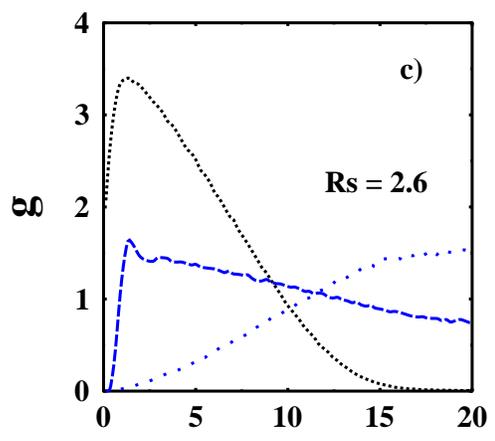
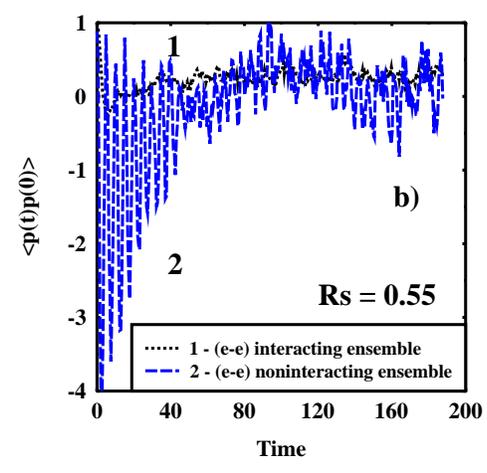
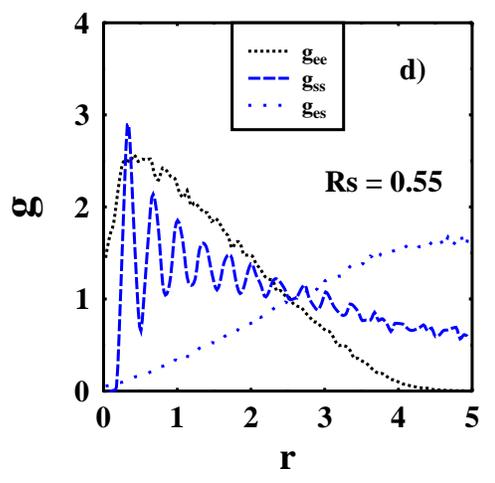
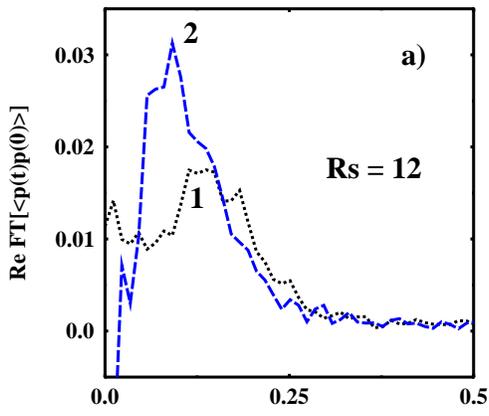
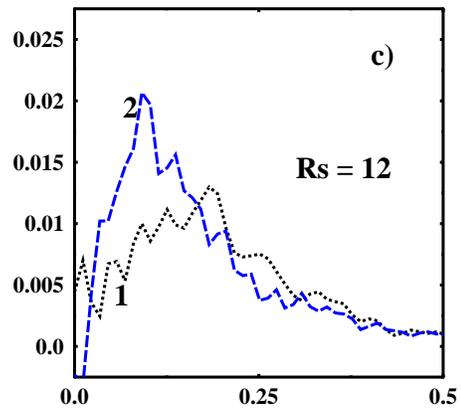
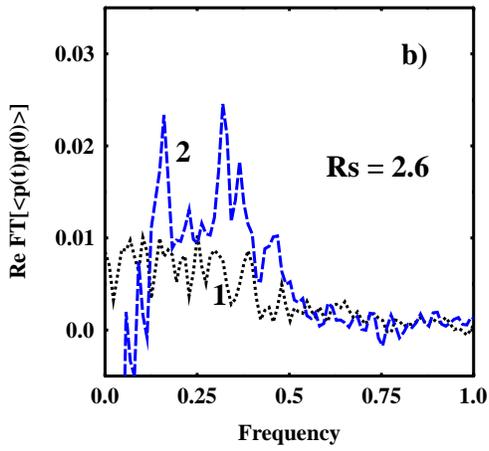
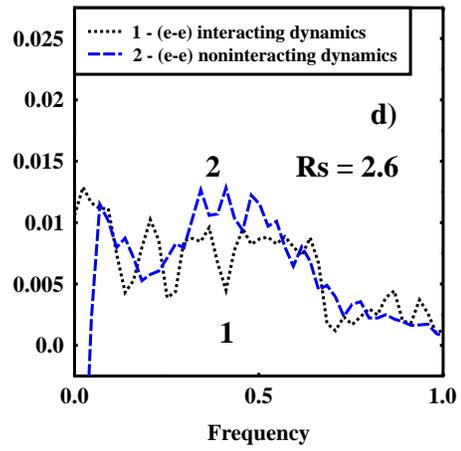
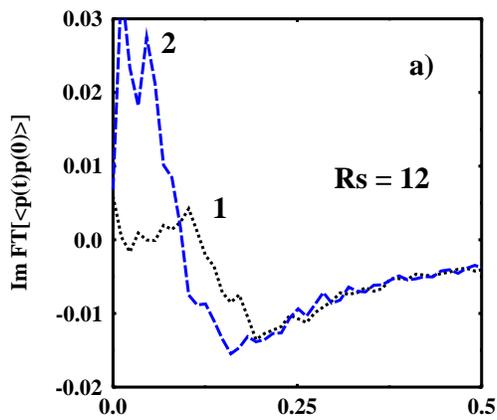
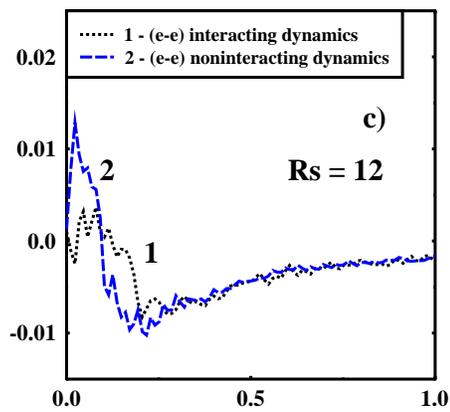
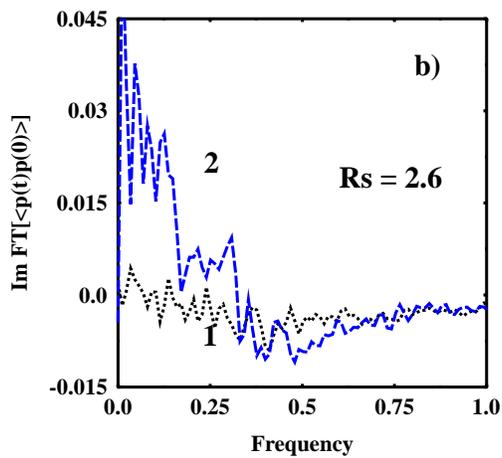
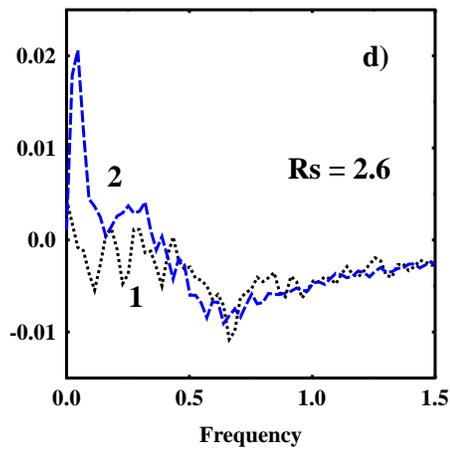
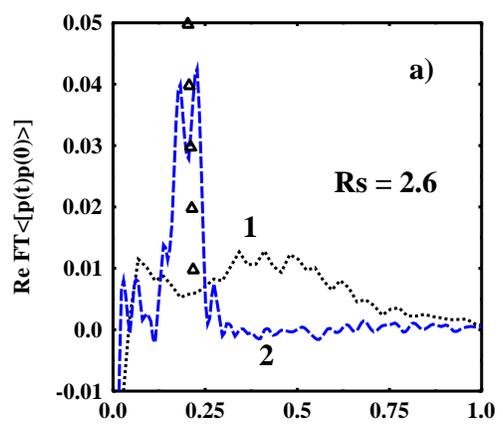
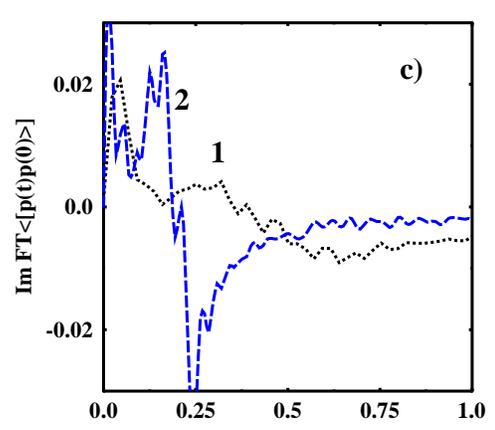
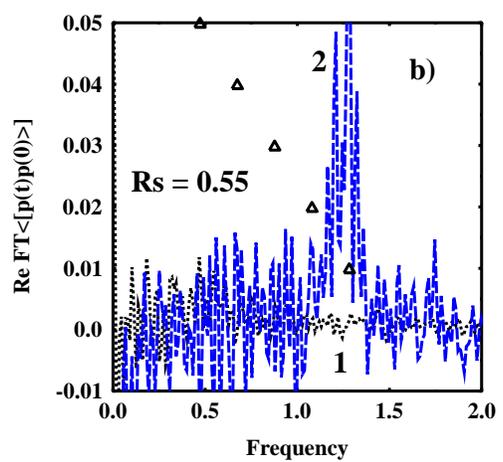
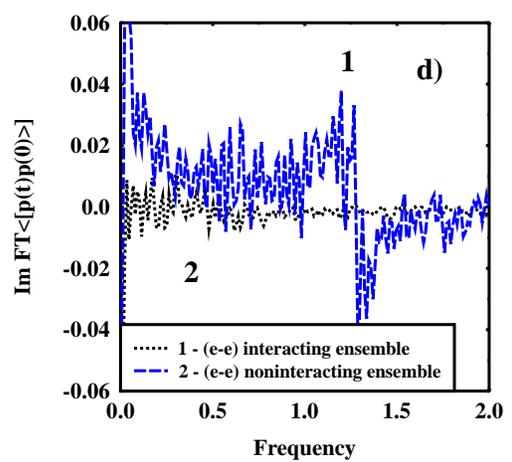
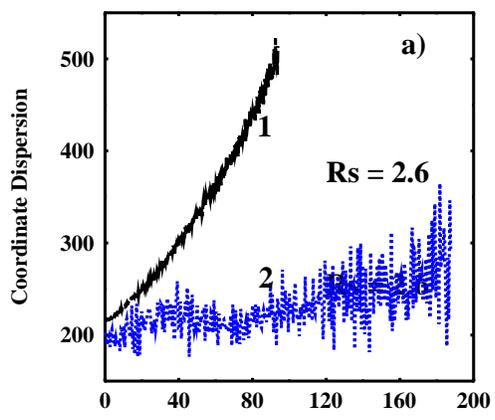
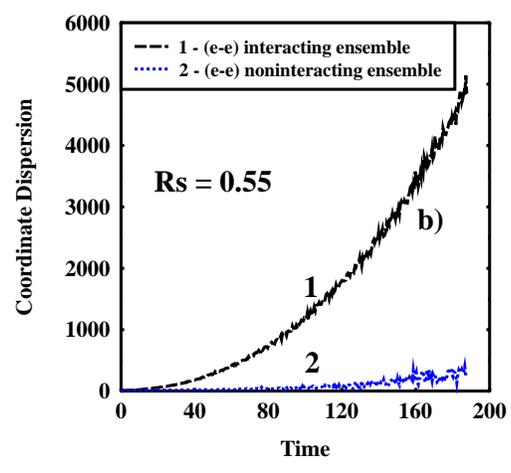
\end{document}